# A Reflective Metasurface for High-Efficiency Terahertz Cross-Polarization Conversion

Muhammad Fayyaz Kashif[1,*], Shobit Agarwal[1], Antonio Iodice[1], Daniele Riccio[1], Junaid Yaseen[2], Zahra Mazaheri[2], Gian Paolo Papari[2], Antonello Andreone[2]

[1]Department of Electrical Engineering and Information Technologies, University of Naples Federico II, 80125 Naples, Italy

[2]Department of Physics "Ettore Pancini", University of Naples Federico II, 80126 Naples, Italy

*muhammadfayyaz.kashif@unina.it

## Abstract

We demonstrate a reflective terahertz (THz) polarization conversion metasurface based on a double-ring aluminum resonator on a quartz substrate. Full-wave simulations are used to optimize the unit-cell geometry and investigate the polarization conversion mechanism through phase analysis, polarization ellipses, and surface-current distributions. The proposed metasurface exhibits near-unity polarization conversion ratio (PCR) at resonance frequencies of 0.333 THz and 0.361 THz. Reflection-mode terahertz time-domain spectroscopy (THz-TDS) measurements show good agreement with the simulated results. The measured PCR exceeds 87% at both resonance frequencies, demonstrating efficient reflection-mode polarization conversion.

## 1. Introduction

Terahertz (THz) radiation provides distinctive capabilities for spectroscopy, imaging, sensing, and high-data-rate wireless communication because its frequencies overlap with molecular rotations, lattice vibrations, collective excitations, and carrier dynamics in a broad range of materials [1], [2]. In these applications, polarization is often an essential degree of freedom rather than a secondary field property. Controlled polarization states can improve spectroscopic selectivity, enhance image contrast, probe anisotropic responses, suppress unwanted backgrounds, and provide additional channels for communication and information encoding.

Conventional THz polarization components are generally based on wire-grid structures, birefringent crystals, multilayer wave plates, or liquid-crystal devices. Although these approaches can provide accurate polarization control, they may require relatively long propagation paths, carefully aligned elements, or materials with suitable birefringence and sufficiently low loss. Their dimensions and dispersive response can also complicate integration into compact quasi-optical systems. Metasurfaces offer an alternative route by using subwavelength resonators to engineer the amplitude and phase of orthogonal field components within a deeply subwavelength thickness [3], [4].

Polarization-conversion metasurfaces can be designed for linear-to-linear, linear-to-circular, circular-to-linear, or more complex multifunctional transformations [5]. In a reflective linear-to-linear converter, an anisotropic patterned layer is commonly combined with a dielectric spacer and a metallic ground plane. The ground plane suppresses transmission, while resonant coupling within the patterned layer and between the resonator and its image currents controls the co- and cross-polarized reflected fields. Efficient conversion is obtained when the co-polarized component is suppressed and the cross-polarized component dominates. The polarization conversion ratio (PCR) is therefore widely used as the principal performance metric. Depending on the number and spectral separation of the excited modes, the same general architecture can support single-band, multiband, or broadband operation.

The development of THz polarization-conversion metasurfaces has progressed from early demonstrations of ultrathin polarization rotation to broadband and multifunctional reflective devices. Grady et al. established a seminal experimental platform for efficient linear polarization conversion and anomalous refraction [5], while Cheng et al. demonstrated how multiple neighboring resonances can be combined in a ground-backed architecture to broaden the conversion band [6]. Subsequent studies explored resonance hybridization, near-field coupling, and increasingly elaborate unit-cell geometries to extend bandwidth or introduce additional polarization functions[7], [8], [9], [10], [11]. These works collectively show that high conversion efficiency can be achieved through several different physical and geometrical strategies, although bandwidth, structural complexity, angular stability, and fabrication requirements remain closely interdependent.

Devices operating in the sub-THz and lower-THz regions are especially relevant to the present work. Kuznetsov et al. experimentally investigated planar bilayer metastructures for sub-THz polarization conversion [12]. Yang et al. later demonstrated a high-efficiency converter based on spiral split-ring resonators, with a resonance close to 0.37 THz [13], whereas Xu et al. used a double-split-ring metasurface to obtain broadband THz polarization conversion [14]. Related reflective designs based on coupled resonators have also been proposed for broader THz bands [8], while wideband and angle-tolerant concepts developed at microwave frequencies provide useful design principles that can be translated toward higher-frequency implementations [15]. Among these examples, structures based on split-ring or double-ring resonators are particularly pertinent because their electric and magnetic responses can be adjusted through a limited number of geometrical parameters and implemented using conventional planar microfabrication.

The broader literature also includes asymmetric metasurfaces supporting multiband and multifunctional operation [10], ultra-broadband reflective concepts at higher THz frequencies [13], and reflective linear-to-circular converters [9]. Such developments underline the versatility of metasurface-based polarization control, but they also make direct comparisons difficult: devices may differ substantially in conversion type, operating frequency, bandwidth definition, number of patterned layers, incidence-angle tolerance, and whether their performance has been verified experimentally.

Active and reconfigurable implementations represent a complementary development path. Recent studies from our group have investigated $VO_2$-integrated reflective metasurfaces for thermally switchable broadband linear-to-linear conversion [16] and for phase-transition-controlled switching between half-wave-plate and quarter-wave-plate responses within a common sub-THz band [17]. These results highlight the potential of phase-change materials for adaptive polarization control, while passive converters remain attractive when structural simplicity, fabrication robustness, and experimentally verified performance are primary requirements.

Experimental demonstrations of simple, passive, reflection-type linear-to-linear converters in the sub-THz region are therefore of particular interest, especially when the measured polarization response can be directly compared with full-wave electromagnetic simulations. In this context, the present study addresses the experimentally demanding realization of a low-complexity reflective polarization converter in the 0.3–0.4 THz range through a complete design–fabrication–characterization workflow.

The proposed double-ring topology supports two neighboring polarization-conversion resonances associated with distinct current configurations, while the influence of the quartz-spacer thickness is explicitly quantified as a fabrication-relevant parameter governing the spectral position of the conversion response. The combination of a conventional aluminum-on-quartz architecture, in-house microfabrication, and polarization-resolved reflection-mode THz-TDS therefore distinguishes the present work from purely numerical demonstrations and provides an experimentally grounded platform for compact sub-THz polarization control.

The remainder of this paper is organized as follows. Section 2.1 presents the metasurface design, numerical modelling, polarization-conversion formalism, resonance analysis, and fabrication procedure. Section 2.2 describes the reflection-mode THz time-domain spectroscopy setup and the comparison between the measured and simulated polarization-resolved responses. Finally, Section 3 summarizes the main conclusions.

## 2. Results and Discussions

### 2.1 Metasurface Design and Fabrication

Figure 1(a) illustrates the three-dimensional schematic of the proposed polarization conversion metasurface, which consists of a periodic array of aluminum double-ring resonators patterned on the top surface of a quartz substrate. The optimized geometrical parameters (see the inset of Fig. 1(a)) are the following: lattice period $P = 333\ \mu$m, outer radius $R_1 = 146\ \mu$m, inner radius $R_2 = 62.5\ \mu$m, inner and outer rings width $w_1 = 42\ \mu$m, central strip width $w_2 = 33\ \mu$m, upper gap $g_1 = 42\ \mu$m, lower gap $g_2 = 100\ \mu$m, quartz substrate thickness $h_{\text{sub}} = 500\ \mu$m, and aluminum thickness $t_{\text{Al}} = 300$ nm. The metallic layer at the bottom acts as a mirror. Under normal incidence of a linearly polarized terahertz wave, the ground plane blocks transmission, resulting in the incident energy to be reflected. The interaction between the incident electromagnetic wave and the patterned aluminum resonator excites localized resonant surface currents, while electromagnetic coupling with the metallic ground plane modifies the complex reflection coefficient, enabling frequency-dependent amplitude and phase modulation of the reflected wave. By properly optimizing the resonator geometry, the metasurface achieves efficient polarization conversion at the designed resonance frequencies.

Full-wave electromagnetic simulations were carried out to optimize the proposed metasurface and evaluate its polarization conversion characteristics. Periodic boundary conditions were applied along the $x$- and $y$-directions, while the structure was excited by a normally incident Floquet port. Quartz was modeled with a frequency-independent relative permittivity of $\varepsilon_r = 3.78$, and the 300 nm thick aluminum layers were modeled as lossy conductors with an electrical conductivity of $3.56\times10^7$ S/m.

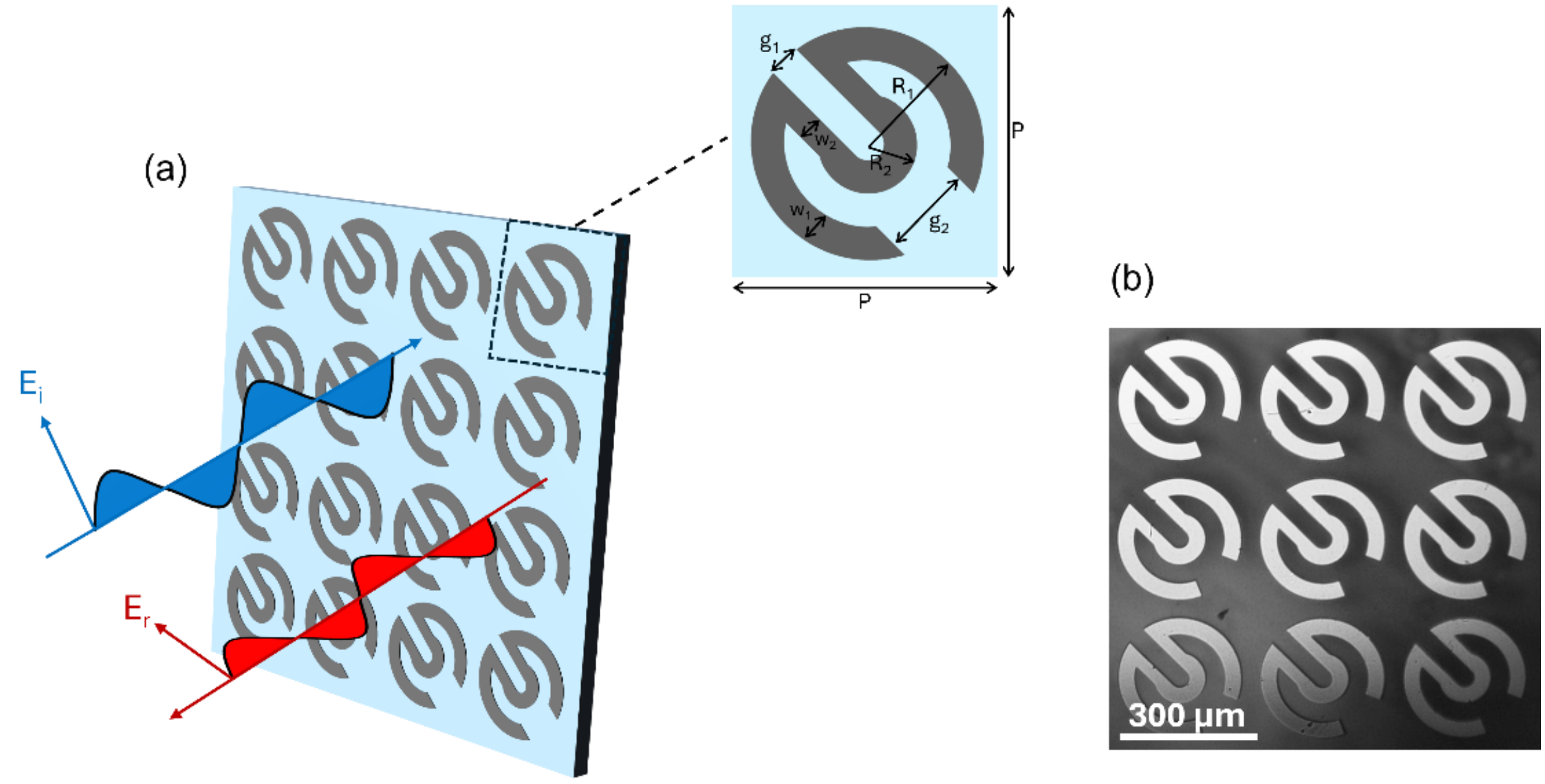

**Figure 1.** Proposed reflective polarization conversion metasurface (PCM) and fabricated device. (a) Three-dimensional schematic of the reflective metasurface illuminated by a normally incident THz (x- or y-polarised) wave. The inset shows the enlarged view of the unit cell showing the geometrical parameters (b) Picture of the fabricated aluminum metasurface on the quartz substrate taken using an optical microscope (magnification 10x).

The relationship between the reflected and incident waves for a plane wave propagating along $z$-direction is described using the Jones matrix theory [18]. Assuming $\vec{E}_r$ and $\vec{E}_i$ are the reflected and incident field vectors, the relationship is given by:

$$\vec{E}_r = J.\vec{E}_i \qquad \text{where} \qquad J = \begin{pmatrix} r_{xx} & r_{xy} \\ r_{yx} & r_{yy} \end{pmatrix} \tag{1}$$

The incident wave can be either $x$- or $y$-linearly polarized. Here $r_{xx}$ and $r_{yy}$ are the co-polarized reflection coefficients while $r_{xy}$ and $r_{yx}$ are the cross-polarized reflection coefficients. Polarization conversion ratio (PCR) is one figure of merit to quantify the efficiency of a linear-to-linear polarization converter. Assuming a y-polarized input wave incidence, the PCR is given by the following formula:

$$PCR = \frac{|r_{xy}|^2}{|r_{xy}|^2 + |r_{yy}|^2} \tag{2}$$

The designed metasurafce can convert both $x$- and $y$-linearly polarized waves to corresponding orthogonally polarized waves. Under $y$-polarized normal incidence, the simulated co- and cross-polarized reflection coefficients are presented in Fig. 2(a). The co-polarized reflection coefficient exhibits pronounced minima at the resonance frequencies of 0.333 THz and 0.361 THz, accompanied by corresponding maxima in the cross-polarized reflection coefficient, confirming conversion of the incident polarization into its orthogonal state. The corresponding polarization conversion ratio (PCR), calculated using Eq. (2), is shown in Fig. 2(b). The metasurface achieves nearly unity PCR values at resonance frequencies. To account for the measured thickness of the commercial quartz substrate, Fig. 2(c) shows the effect of varying the substrate thickness from 460 to 500 µm. As the substrate thickness increases, the PCR band gradually shifts toward lower frequencies. This shift occurs because a thicker quartz spacer increases the optical path length, leading to greater phase accumulation of the wave reflected from the metallic ground plane. The resulting change in the interference condition between the reflected wave and the resonator shifts the polarization conversion band to lower frequencies [19].

To further understand the polarization conversion mechanism, the phase difference between the co- and cross-polarized reflected fields for the metasurface with a 500 μm substrate is shown in Fig. 3(a). At both resonance frequencies 0.333 and 0.361 THz, the relative phase difference ($\Delta\varphi = \varphi_{yy} - \varphi_{xy}$) approaches 180º, while the cross-polarized reflection dominates over the co-polarized component. This combination of amplitude and phase characteristics satisfies the condition for efficient linear-to-linear polarization conversion. The polarization states of the reflected wave at the two resonance frequencies are illustrated by the polarization ellipses in Fig. 3(b). Since the co-polarized reflection is strongly suppressed while the cross-polarized component remains dominant, the reflected field becomes nearly linearly polarized along the orthogonal direction, confirming efficient polarization rotation by the proposed metasurface.

To elucidate the underlying resonance mechanism, the simulated surface current distributions on the patterned resonator and the metallic ground plane at the two resonant frequencies are presented in Fig. 3(c). At 0.333 THz, the surface currents on the resonator and the induced image currents on the ground plane flow predominantly in opposite directions, indicating the formation of a magnetic-dipole resonance through strong resonator-ground coupling. In contrast, at 0.361 THz, the current distribution becomes more complex and partially localized over different sections of the resonator, while the corresponding ground-plane currents are less strictly anti-parallel. This behavior suggests the excitation of a hybrid electric–magnetic resonant mode. These distinct resonant mechanisms are responsible for the two high-efficiency polarization conversion bands observed in the proposed metasurface.

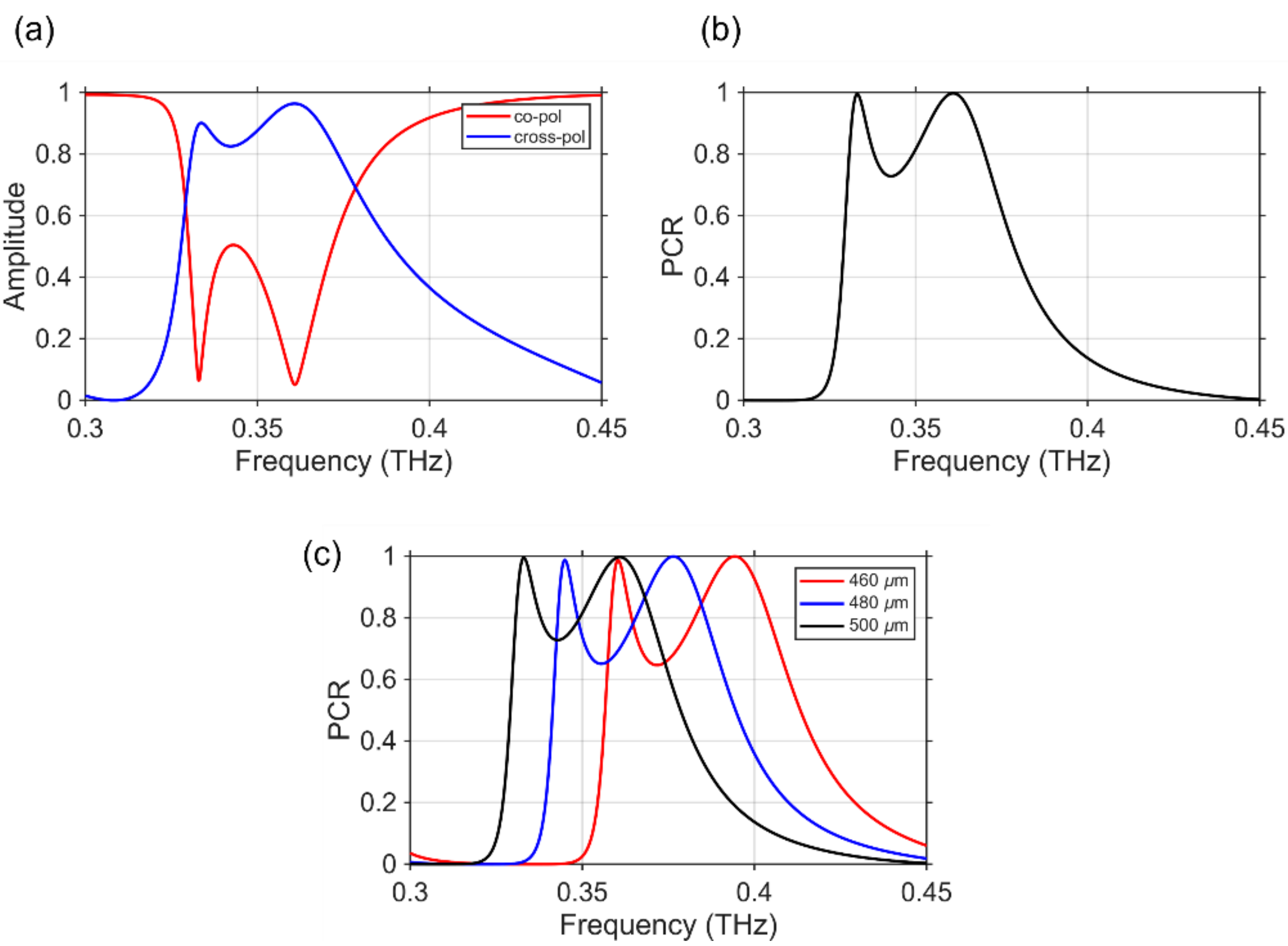

Figure 2. Numerical analysis of the proposed reflective metasurface. (a) Co-polarized reflection coefficients, (b) cross-polarized reflection coefficients, and (c) polarization conversion ratio (PCR) for different quartz substrate thicknesses of 460 $\mu$m, 480 $\mu$m and 500 $\mu$m.

The proposed metasurface was fabricated on a 1×1 $cm^2$ quartz substrate using a standard microfabrication procedure in the FABRIS facility of the University of Naples "Federico II". First, a 300 nm thick aluminum film was deposited on both sides of the quartz substrate by physical vapor deposition (PVD) sputtering to form the patterned resonator and the reflective ground plane. Subsequently, maskless photolithography was employed to define the metasurface pattern. A positive photoresist (S1813) was spin-coated at 6000 rpm for 1 min, followed by soft baking at 80 ºC for 20 min and pattern exposure using a laser writer. The exposed photoresist was developed in MF-319 developer to transfer the designed pattern onto the aluminum layer. Finally, the unprotected aluminum was removed by wet chemical etching using an aluminum etchant, while the remaining photoresist was stripped to obtain the fabricated metasurface. An optical microscope image of the fabricated device is shown in Fig. 1(b).

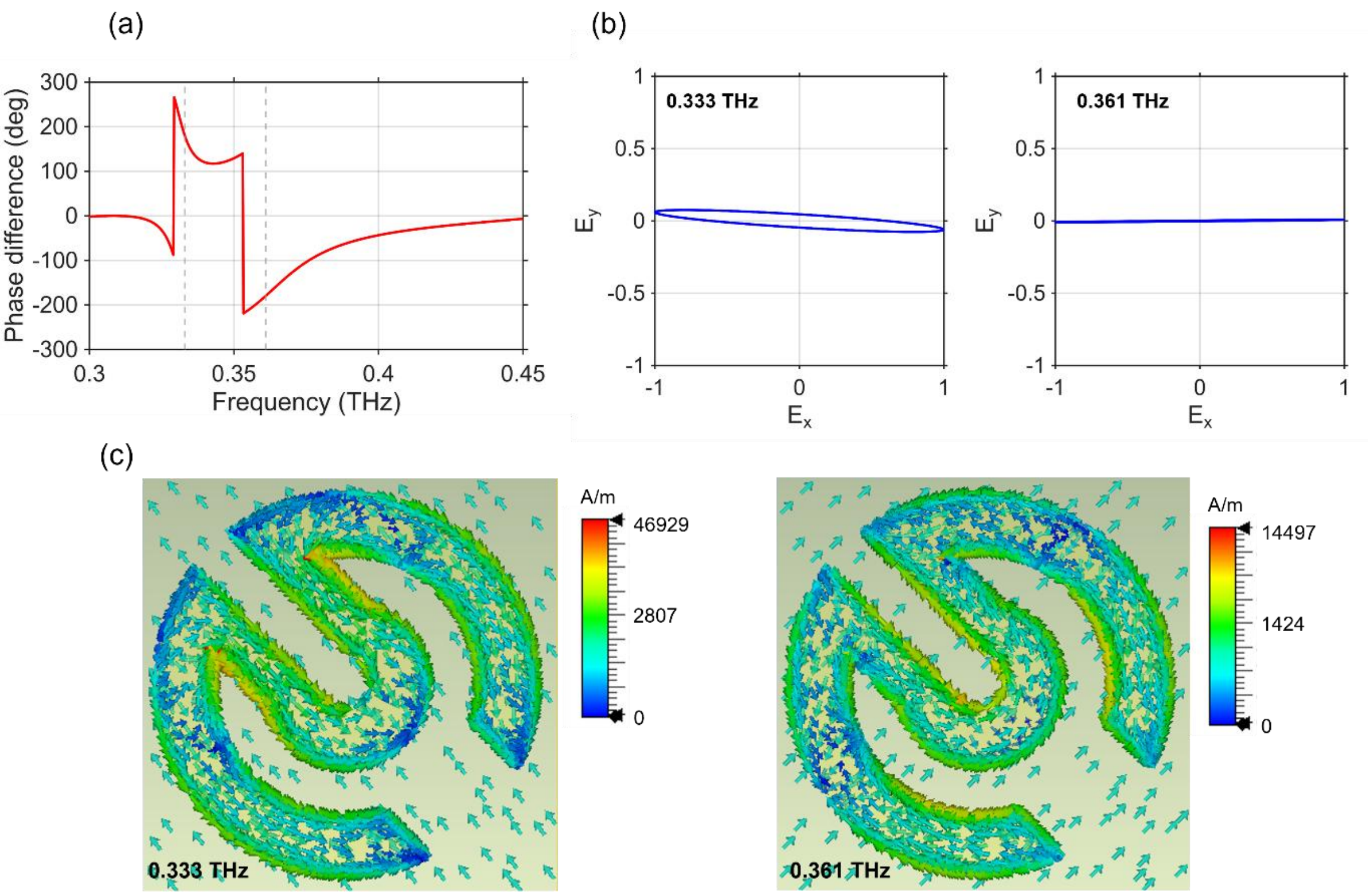

Figure 3. (a) Phase difference between the co- and cross-polarized reflected field components for the 500 $\mu$m substrate thickness. The dashed vertical lines at resonance frequencies of 0.33 and 0.361 THz correspond to 180° phase difference. (b) Polarization ellipses at 0.333 and 0.361 THz. (c) Surface-current distributions on the patterned resonator and corresponding image currents on ground plane at the two resonance frequencies.

## 2.2 Metasurface THz-TDS Characterization

The fabricated metasurface was characterized using a terahertz setup, as illustrated in Fig. 4(a). The setup is based on a commercial time domain spectrometer (TDS), where a 1550 nm femtosecond fiber laser generates an optical pulse train which is delivered to two photoconductive antennas (PCAs) through optical fibers (TERAK15, Menlo Systems). THz pulses are generated at the biased emitter antenna, while the detector antenna samples the incoming THz transient as a function of the delay between the two optical pulses. Reconstruction of the THz transient is ensured by a motorized delay line in the emitter arm, which controls the relative arrival time between the excitation and sampling pulses.

In the reflection-mode, a broadband linearly polarized terahertz pulse is directed onto the metasurface under normal incidence, while a flat metallic mirror is used as the reference reflector. Since the detector is sensitive to only one polarization component, the co- and cross-polarized reflected fields were measured using two wire-grid-polarizers with a high extinction ratio (> 40 dB in the frequency range of interest), following the method reported in [20]. The first polarizer was oriented at 45º with respect to the detector polarization axis, while the second polarizer was aligned either parallel (45º) or orthogonal (135º) to the first polarizer to measure the co-polarized and cross-polarized reflection components, respectively. This configuration enables accurate extraction of the orthogonal polarization components of the reflected terahertz field.

The measured time-domain waveforms of the reference, co-polarized, and cross-polarized reflected signals are presented in Fig. 4(b). Fourier transformation of these signals yields the corresponding frequency-domain reflection coefficients shown in Fig. 4(c) and Fig. 4(d). The co-polarized reflection coefficient exhibits two pronounced minima at 0.360 THz and 0.394 THz, while the cross-polarized reflection reaches its maximum values at the same frequencies, confirming efficient polarization conversion. The measured spectra agree well with the simulated results, with small deviations in amplitude and resonance position. These discrepancies are primarily attributed to fabrication tolerances, material parameters, and experimental uncertainties associated with the THz-TDS alignment and calibration. The polarization conversion ratio, calculated from the measured reflection coefficients using Eq. (2), is compared with the simulated results in Fig. 4(e). The experimental PCR closely follows the simulated response, reaching values above 87% at both resonance frequencies, thereby experimentally validating the high polarization conversion efficiency of the proposed metasurface.

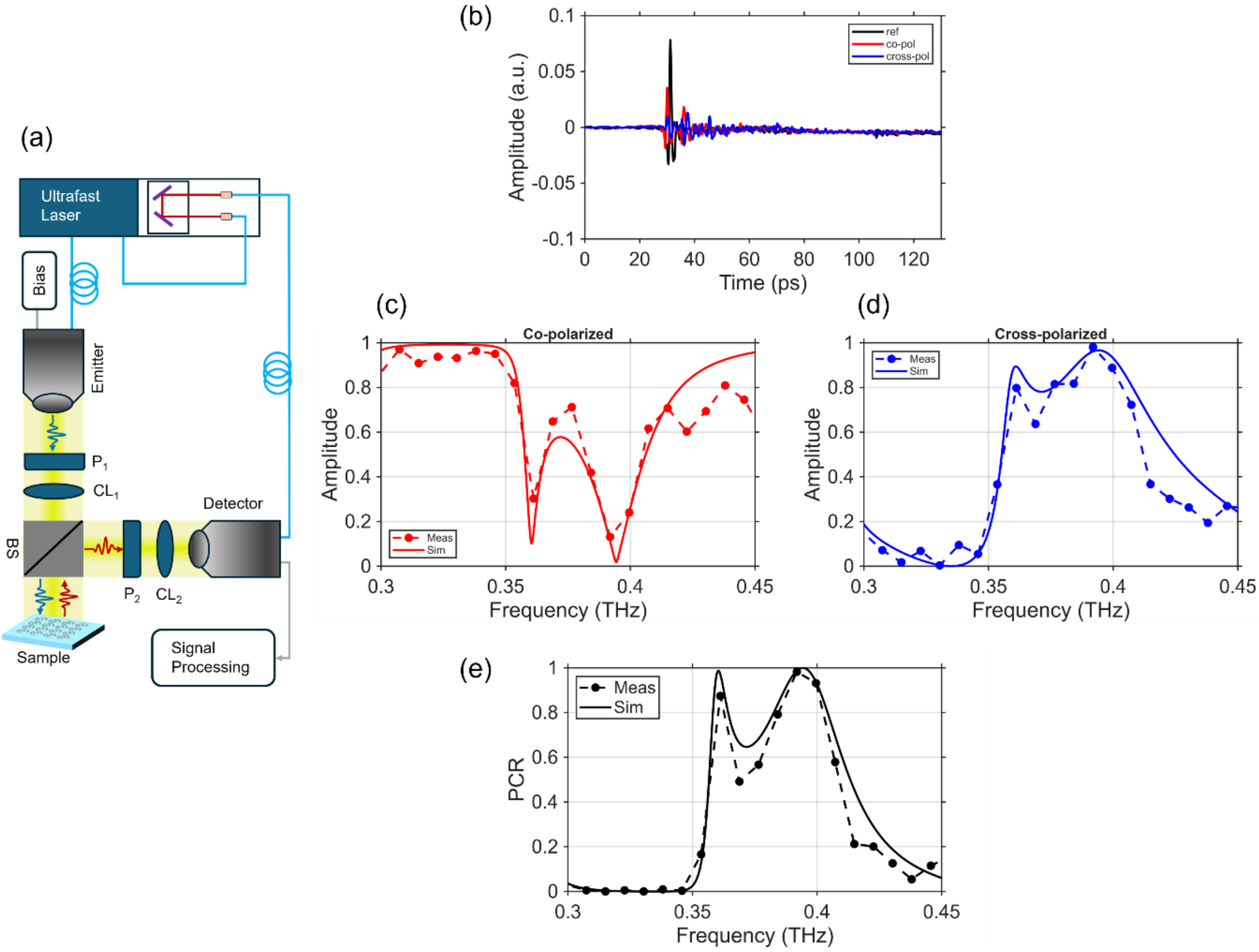

Figure 4. (a) Schematic of terahertz time-domain spectroscopy (THz-TDS) experimental setup operated in reflection mode for characterization of the fabricated metasurface. (b) The measured time-domain THz-TDS signals of the reference pulse as well

as the reflected co- and cross-polarized components. (c) Measured (dashed) and simulated (solid) reflectance spectra of the co-polarized component under normal incidence. (d) Measured (dashed) and simulated (solid) reflectance spectra of the cross-polarized component under normal incidence (e) Measured (dashed) and calculated through simulations (solid) PCR.

## 3. Conclusion

We demonstrate a reflective terahertz polarization conversion metasurface based on a double-ring aluminum resonator on a quartz substrate. The proposed metasurface achieves near-unity polarization conversion ratio in simulations, while the fabricated device exhibits PCR values exceeding 87% at resonance frequencies of 0.360 THz and 0.394 THz. Reflection-mode terahertz time-domain spectroscopy measurements show good agreement with the simulated results. Phase analysis, polarization ellipses, and surface-current distributions provide physical insight into the polarization conversion mechanism. The proposed metasurface offers a compact and experimentally validated solution for reflection-mode polarization control for terahertz waves.

Beyond the achieved polarization-conversion values, the main contribution of this work is the experimental closure of the complete design–fabrication–validation loop for a low-complexity sub-THz metasurface. The demonstrated aluminum-on-quartz architecture thus provides a practical baseline for future polarization-control devices incorporating active materials, electrically or thermally tunable elements, or spatially varying unit cells, while retaining the same microfabrication and THz-TDS characterization framework.

Future developments may extend this passive architecture toward actively tunable or switchable operation through the integration of phase-change or other reconfigurable materials, while preserving its simple ground-backed configuration. Further investigations of angular stability, fabrication tolerances, and array-level integration will be important for assessing its use in adaptive sub-THz components and communication front ends.

## Funding

This work was supported by the European Union under the Italian National Recovery and Resilience Plan (NRRP) of NextGenerationEU, partnership on “Telecommunications of the Future” (PE00000001 - program “RESTART”).